\documentclass[12pt]{iopart}
\usepackage{graphicx}
\usepackage{esint}
\usepackage{subfig}
\usepackage{textcomp}
\usepackage{braket}

\begin{document}

\title[]{Quantum Uncertainty in the Beam Width of Spatial Optical Modes}

\author{Vanessa Chille$^{1,2,3}$, Peter Banzer$^{1,2,4}$, Andrea Aiello$^{1,2}$, \newline Gerd Leuchs$^{1,2,4}$, Christoph Marquardt$^{1,2}$, Nicolas Treps$^{3}$, Claude Fabre$^{3}$}

\address{$^1$Max Planck Institute for the Science of Light, Guenther-Scharowsky-Str. 1/Bldg. 24, D-91058 Erlangen, Germany}
\address{$^2$Institute of Optics, Information and Photonics, University of Erlangen-Nuremberg, Staudtstr. 7/B2, D-91058 Erlangen, Germany}
\address{$^3$Laboratoire Kastler Brossel, Sorbonne Universit\'e - UPMC, ENS, Coll\`ege de France, CNRS; 4 place Jussieu, 75252 Paris, France}
\address{$^4$Department of Physics, University of Ottawa, 25 Templeton, Ottawa, Ontario, K1N 6N5 Canada}

\ead{vanessa.chille@mpl.mpg.de}

\begin{abstract}
We theoretically investigate the quantum uncertainty in the beam width of transverse optical modes and, for this purpose, define a corresponding quantum operator. Single mode states are studied as well as multimode states with small quantum noise. General relations are derived, and specific examples of different modes and quantum states are examined. For the multimode case, we show that the quantum uncertainty in the beam width can be completely attributed to the amplitude quadrature uncertainty of one specific mode, which is uniquely determined by the field under investigation. This discovery provides us with a strategy for the reduction of the beam width noise by an appropriate choice of the quantum state.
\end{abstract}

\vspace{2pc}

\maketitle

\section{Introduction}
It is well-known that light is always affected by quantum noise that cannot be totally eliminated. This noise is the ultimate limitation to any measurement performed by optical means. By considering more precisely the measurement process \cite{HenryKazarinov} and understanding the underlying effects, however, one can find ways to improve particular measurement precisions by utilizing an individually tailored quantum state. For instance, absorption measurements have been improved below the standard quantum limit by using sub-Poissonian light \cite{Davidovich1996, Polzik, Marin}, interferometric measurements by utilizing vacuum squeezed light \cite{Xiao, Grangier, LIGO} and differential measurements by using twin beams \cite{Gao, Schwob}.\newline
More recently, studies have been devoted to the spatial properties of quantum light \cite{ReviewKolobov, Gabriel}. In \cite{FrankeArnold2004}, for example, the experimental observation of an uncertainty principle of angular position and momentum has been reported. Different groups have considered the issue of measuring spatial properties of light beams below the standard quantum limit. The first parameter that has been considered is the accurate determination of the center of a light beam, and the measurement of the smallest possible deflection \cite{Shin, Treps2002}. \cite{QuLaserPointer} presented an improvement of the measurement sensitivity for a beam's pointing direction by an appropriate choice of the quantum state used in the measurement, and in particular by optimizing the shape of the transverse modes in which the quantum state is defined. Although being a topic of on-going research, certain aspects of the spatial properties of nonclassical light deserve more attention than they have received so far. To the best of our knowledge, the quantum noise in the width of a beam's cross section has never been studied before. 
In many different applications in particular the resolution capabilities depend on the spot size of a focussed beam, such as optical trapping \cite{TrappingBowman}, lithography \cite{LithographyBrueck}, confocal microscopy and optical data storage. The spot size can be reduced by taking advantage of classical properties, for example by a clever choice of the spatial and vectorial character of the utilized modes \cite{Quabis}. But the noise in the spot size is still an open question. As the uncertainty in the beam width has not even been studied for the paraxial case yet, and the width of a paraxial beam is a fundamental beam parameter, we will address this problem in the present article. \newline
The first question that arises, is how to actually define the beam width in both classical and quantum scenarios. The natural second question to follow is the one about the uncertainty in this beam width and its fundamental quantum limits. 
To the first question, there is more than one answer. Many different measures exist that are used according to what is most appropriate for the specific problem \cite{TutorialSiegman}. A very common choice is, for example, to define a beam's width as the distance between the points, at which the intensity has dropped down to $1/e^2$ or $1/2$ (FWHM) of the maximum value. In particular for high power lasers, one may also define the beam width according to how much power is illuminating a particular surface. 
If one is interested in theoretical descriptions of the distribution under investigation, one very often chooses to work with the standard deviation as a measure of the beam width. We decided to use a closely related definition and use the spatial variance of the intensity distribution as a measure for the beam width.\newline
Using this measure, the aforementioned second question about quantum noise corresponds to studying the uncertainty of the variance of the spatial intensity distribution or, equivalently, to investigating the second moment of the intensity distribution \cite{Alda}. This has, to the best of our knowledge not been studied in literature so far.\newline
Here, we start with working on the topic of noise in the second moment by investigating the fundamental quantum limits for paraxial beams. We proceed as follows: First, we present the definition of an operator measuring the beam width and derive its variance. This gives an expression for the uncertainty in the beam width that is used for studying the cases of single mode and multimode states. For the single mode case, examples for different modes and quantum states are investigated. In particular, we show that quantum states can indeed be chosen such that the the quantum noise in the beam width is reduced. For the multimode case, we construct the so-called detection mode. The noise in the beam width can be attributed completely to the noise in the amplitude quadrature of this mode. Finally, we discuss the conjugate variable of the beam width.\newline
\section{Definition of the Beam Width}
\begin{figure}[h]
\centering
    \includegraphics[width=0.5\textwidth]{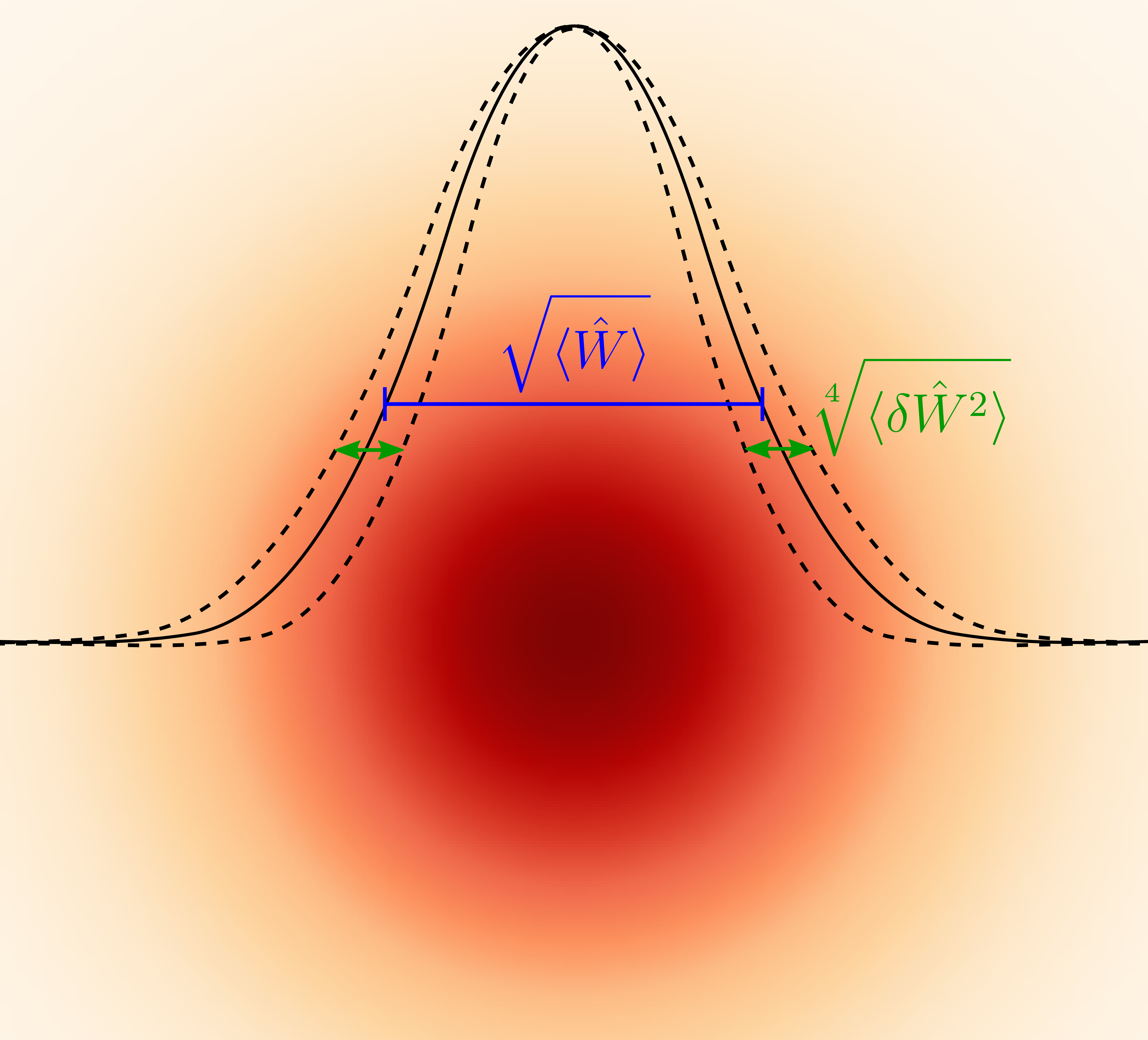}
		\caption{The intensity distribution in the cross section of a fundamental Gaussian beam is depicted. $\langle \hat{W}\rangle$ represents the variance of the transverse distribution (see Eq.\,\ref{EqBeamWidthOperator}), the indicated quantity $\sqrt{\langle \hat{W}\rangle}$ is thus its standard deviation. The uncertainty in the width of the distribution can be quantified by $\langle \delta\hat{W}^2 \rangle$ (see Eq.\,\ref{varvar}). In the figure, $\sqrt[4]{\langle \delta\hat{W}^2 \rangle}$ is indicated, possessing the same dimension as $\sqrt{\langle \hat{W}\rangle}$.}
    \label{fig:sketch}
\end{figure}

As discussed above, we define the width of a beam of light as the spatial variance of the transverse intensity distribution. For further investigations, we assume a propagation of the light beam along the $z$ direction and study its transverse shape. We expand the light field $\hat{E}^{(+)}(x,y,z)$ describing the beam in the mode basis $\{u_i(x,y,z)\}$. $\{u_i(x,y,z)\}$ is an arbitrarily chosen complete and orthonormal basis formed by solutions of the paraxial wave equation. We perform our studies in a fixed plane $z=$ const, as a matter of convenience, we choose this to be $z=0$. We can thus omit the propagation factor and use the transverse part of the light field $\hat{E}^{(+)}(x,y)$, and the transverse mode basis $\{u_i(x,y)\}$. For each transverse mode, a set of creation and annihilation operators $\hat{a}_i$ and $\hat{a}_i^\dagger$ is associated. The transverse light field operator can thus be written as 
$\hat{E}^{(+)}(x,y) = \sum_i u_i(x,y) \hat{a}_i$ \cite{ReviewKolobov}.\newline
We define the operator $\hat{W}$ as the spatial variance of the intensity distribution as follows:
\begin{eqnarray}
\hat{W} &= \frac{1}{\langle \iint \hat{E}^{(-)}(x,y) \hat{E}^{(+)}(x,y) \mathrm{d}x\mathrm{d}y \rangle} \iint (x^2+y^2) \hat{E}^{(-)}(x,y) \hat{E}^{(+)}(x,y) \mathrm{d}x\mathrm{d}y\nonumber\\
 &= \frac{1}{N_\mathrm{all}} \sum_{i,j} D_{ij} \hspace{0.1cm} \hat{a}_i^\dagger \hat{a}_j, \label{EqBeamWidthOperator}
\end{eqnarray}
with
\begin{eqnarray}
D_{ij} &= \iint (x^2+y^2) u_i^*(x,y) u_j(x,y) \mathrm{d}x\mathrm{d}y \label{EqDij}
\end{eqnarray}
and
\begin{eqnarray}
N_\mathrm{all} &= \sum_i N_i = \sum_i \langle\hat{a}_i^\dagger \hat{a}_i\rangle.
\end{eqnarray}
Fig.\,\ref{fig:sketch} indicates the physical meaning of the mean value $\langle \hat{W} \rangle$ for a fundamental Gaussian beam.\newline
Our definition of the beam width is very appropriate for stigmatic modes that are symmetric in $x$ and $y$ directions, as one averages over the transverse directions $x$ and $y$. For astigmatic modes, however, a measure taking into account the noise along the different cross-sectional axes individually would be more meaningful. In this paper, we will therefore focus on stigmatic modes.\newline
Defining the beam width as the variance of the spatial intensity distribution is a legitimate choice, but, of course, other measures would be equally justified. Our approach is also applicable if we replace $(x^2+y^2)$ by any different measure $f(x,y)$. As a matter of convenience, we use $(x^2+y^2)$ in the whole course of the paper. The variance has, of course, the dimension of an area, not of a length. Nevertheless, the variance represents a measure for the beam width.
\section{Noise in the Beam Width} \label{SecGenBWN}
In order to find an appropriate expression for the quantum uncertainty in the beam width, we determine the variance of the beam width operator $\langle \delta \hat{W}^2 \rangle$, with $\delta \hat{W} = \hat{W}-\langle \hat{W} \rangle$.\newline
After a straightforward calculation, we find
\begin{eqnarray}\label{varvar}
\langle \delta \hat{W}^2 \rangle = \frac{1}{N_\mathrm{all}^2}\left[\sum_{ijkl} D_{ij} D_{kl} \left(\langle \hat{a}_i^\dagger \hat{a}_k^\dagger \hat{a}_j \hat{a}_l \rangle - \langle \hat{a}_i^\dagger \hat{a}_j \rangle \langle \hat{a}_k^\dagger \hat{a}_l \rangle\right) + \sum_{il} F_{il} \langle \hat{a}_i^\dagger \hat{a}_l \rangle \right],
\end{eqnarray}
with
\begin{eqnarray}
F_{il} = \iint (x^2+y^2)^2 u_i^*(x,y)u_l(x,y)\mathrm{d}x\mathrm{d}y. \label{EqFil}
\end{eqnarray}
In the measurement process, the quantum uncertainty manifests itself in the form of noise. We may thus refer to $\langle \delta \hat{W}^2 \rangle$ also as 'noise', meaning the measured noise due to the fundamental property of quantum uncertainty.\newline
Fig.\,\ref{fig:sketch} visualizes the quantities under investigation for the example of a fundamental Gaussian beam. 
Eq.\,\ref{varvar} shows that the uncertainty in the beam width depends, on the one hand, on the spatial distribution, i.e. the spatial modes that are contained in the integral expressions $D_{ij}$ and $F_{il}$. On the other hand, it depends on the quantum state coming into play due to the expectation values of products of the annihilation and creation operators associated with different modes of the basis. We thus want to explore the dependence of both, the spatial mode and the quantum state, on the beam width uncertainty. We will do so, first, for the very common case of single mode states and investigate explicit examples. Our second approach that we will consider later treats a more general case: multimode states, but with small quantum noise.\newline
At this point, the question arises why we actually have to distinguish these two cases. Of course, a statistical superposition of modes cannot be expressed as a single mode in any case. But, in the classical case, at least every coherent superposition of light fields is single mode. Even if, at first glance, it seems to be multimode in one basis, it is always possible to define a new basis in which the first mode is the superposition of all the modes of the initial basis that contribute to the light field. Thus, the state is single mode. 
In quantum optics, however, this is not the case. Light fields exist that are multimode although they are not statistical superpositions. One cannot find a basis in which these light fields factorize such that only one of the states is not a vacuum state. A superpostion of transverse modes, for instance, is a quantum multimode state if only one of the modes is a noncoherent state. Detailed proofs and further explanations can be found in \cite{Treps2005}.\newline
\subsection{Single Mode States} \label{Sec:SingleModeStates}
\begin{figure}[h]
    \subfloat[][]{\includegraphics[width=0.5\textwidth]{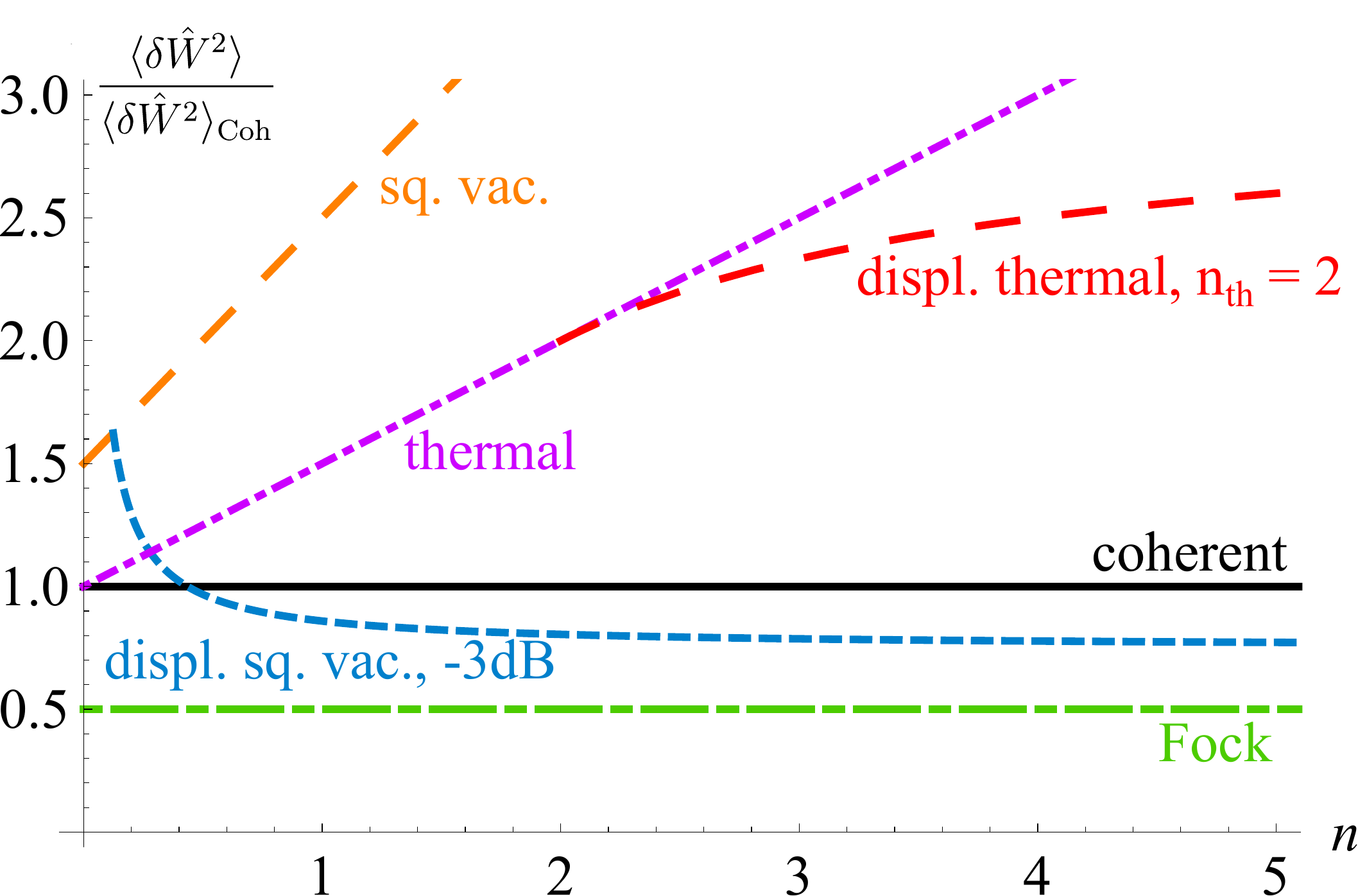}}
   \hspace{0.3cm}
  \subfloat[][]{\includegraphics[width=0.45\textwidth]{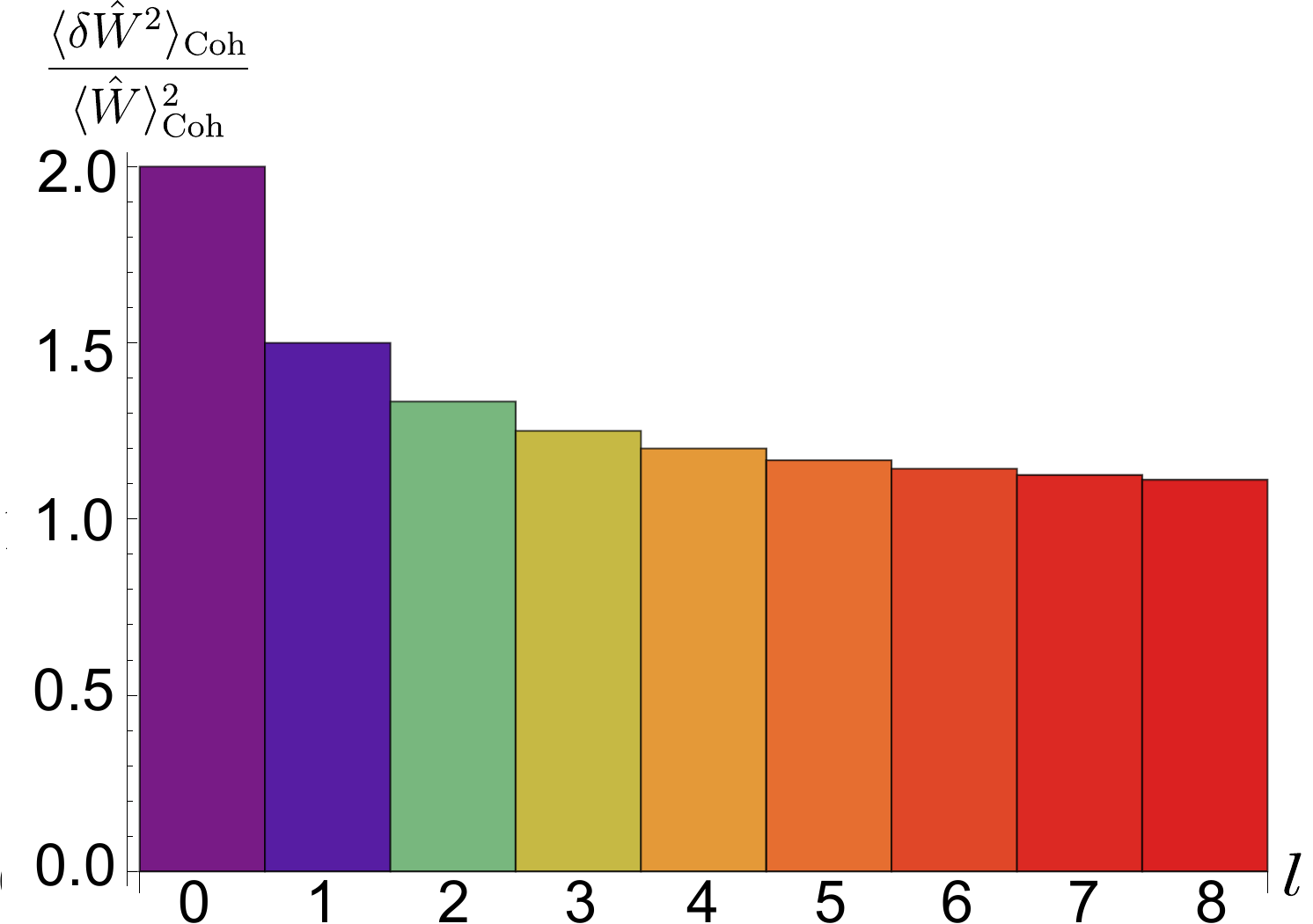}}
    \caption{(a) \textit{Noise in the beam width of a fundamental Gaussian beam for different single mode quantum states, normalized with the noise of the coherent state.}\newline The uncertainty in the beam width depends strongly on the quantum state. The lowest noise is achieved for a Fock state, for which the photon number fluctuations are equal to $0$.\newline (b) \textit{Noise in the beam width of a single mode coherent state in a Laguerre-Gauss mode LG$_{lp}$ with $p=0$ and different values for the azimuthal values of $l$, normalized with the squared mean value of the beam width.}\newline The influence of the quantum noise is getting less significant for larger total beam sizes. The uncertainty of the beam width decreases with increasing azimuthal parameter $l$.}
    \label{fig:SingleMode}
\end{figure}

For single mode states, Eq.\,\ref{varvar} simplifies significantly. Due to the normal ordering, the sums over all basis modes reduce to only one term with the index $i=k=j=l=0$.\\
We thus find
\begin{eqnarray}
\langle \delta\hat{W}^2\rangle = \frac{1}{\langle \hat{n}_0\rangle}\left[D_{00}^2\left(\frac{\langle\delta\hat{n}_0^2\rangle}{\langle\hat{n}_0\rangle}-1\right) + F_{00}\right],
\end{eqnarray}
where the factor $\left(\frac{\langle\delta\hat{n}_0^2\rangle}{\langle\hat{n}_0\rangle}-1\right)$ corresponds to Mandel's Q parameter \cite{MandelQParameter}.
Hence the noise in the beam width depends not only on the mean photon number $\langle\hat{n}_0\rangle$ and the spatial intensity distribution given by the classical mode, but also on the variance of the photon number $\langle\delta\hat{n}_0^2\rangle$. The noise in the spatial distribution of the mode is thus influenced by the quantum properties of the state. In the following, we want to investigate the uncertainty for different spatial distributions and quantum states.\newline
An important question arising in particular in the context of practical implementations, is the normalization. In other words, the definition of a quantity, the noise $\langle \delta\hat{W}^2\rangle$ is compared to. We decided to follow the same approach, which is the standard for quadrature squeezing: we use the uncertainty of a coherent state with the same mean photon number in the same spatial mode as a reference. 
By studying different quantum states in one particular mode, we will thus get the dependence of the noise in the beam width on the quantum state allowing an easy comparison of different quantum states, since the same reference is utilized. If we want to study different spatial modes, it actually tells us how the spatial mode dependence for the quantum state under investigation differs from that of the coherent state. Alternatively, we may instead normalize it with the mean value $\langle\hat{W}\rangle = D_{00}$ which does not depend on the quantum state but only on the spatial mode. This allows us to study the full spatial mode dependence of the noise for the state under investigation relative to the mean beam width. In the following, we will comment on the spatial mode dependence and give examples. In that case, we use $\langle \hat{W} \rangle$ for normalization. But to begin with, we present the dependence of the quantum noise in the beam width for different quantum states and will thus utilize $\langle \delta\hat{W}^2\rangle_\mathrm{Coh}$ as normalization factor. We determine the relative noise of the beam width $\frac{\langle \delta\hat{W}^2\rangle_\mathrm{test}}{\langle \delta\hat{W}^2\rangle_\mathrm{Coh}}$ for different quantum states. The normalization factor is found to be $\langle \delta\hat{W}^2\rangle_\mathrm{Coh} = \frac{F_{00}}{\bar{n}_0}$, with $\bar{n}_0 = |\alpha|^2$ and $|\alpha|$ being the amplitude of the coherent state \cite{Loudon}. In Fig.\,\ref{fig:SingleMode}(a), the results for the case of a fundamental Gaussian mode in different single mode quantum states are visualized that are discussed in the following. 
For a Fock state, we have
\begin{eqnarray}
\frac{\langle \delta\hat{W}^2\rangle_\mathrm{Fock}}{\langle \delta\hat{W}^2\rangle_\mathrm{Coh}} = 1-\frac{D_{00}^2}{F_{00}}.
\end{eqnarray}
The relative noise is thus independent of the mean photon number in this case. For a vacuum squeezed state, we determine 
\begin{eqnarray}
\frac{\langle \delta\hat{W}^2\rangle_\mathrm{SqVac}}{\langle \delta\hat{W}^2\rangle_\mathrm{Coh}} &= \frac{D_{00}^2}{F_{00}} (2\bar{n}_0+1)+1,
\end{eqnarray}
with $\bar{n}_0 = \sinh^2(s)$, where $s$ is the squeezing parameter \cite{Loudon}. Here, $s$ only determines the mean photon number, but does not influence the uncertainty further. As the state is located at the üphase space origin, the phase is not distinct and the noise steming from the antisqueezed parameter results in a relatively high noise. A higher amount of squeezing will then cause only higher noise.\newline\newpage
For a displaced amplitude squeezed state, the squeezing parameter $s$ plays a more elaborate role: 
\begin{eqnarray}
\frac{\langle \delta\hat{W}^2\rangle_\mathrm{DisplSq}}{\langle \delta\hat{W}^2\rangle_\mathrm{Coh}} = &\left[ -\sinh^2(s) e^{-2s} + 2\sinh^2(s)(\sinh^2(s)+1)\right]\frac{D_{00}^2}{F_{00}\bar{n}_0}\nonumber\\&+\left[ 1+\frac{D_{00}^2}{F_{00}}(e^{-2s}-1) \right],
\end{eqnarray}
with $\bar{n_0} = \sinh^2(s) + |\alpha|^2$, where $|\alpha|$ is again the displacement \cite{Loudon}. 
For a given total photon number, one may find an optimal squeezing for which the noise in the beam width is smaller than the one for the coherent state. In contrast, if the displacement is too small, the state is very similar to squeezed vacuum and the noise is very high. If the displacement is too big, we approach the coherent state, with a noise that is still higher than the one of a Fock state. Thus, we have to find the right trade-off between displacement, i.e., the right distance from the origin in phase space, and the amplitude squeezing.\newline
For a thermal state, we arrive at the expression 
\begin{eqnarray}
\frac{\langle \delta\hat{W}^2\rangle_\mathrm{Thermal}}{\langle \delta\hat{W}^2\rangle_\mathrm{Coh}} =\frac{D_{00}^2}{F_{00}}\bar{n}_0+1.
\end{eqnarray}
The mean photon number is given by $\bar{n}_0 = [e^{\beta\hbar\omega}-1]^{-1}$, with $\beta = (k_\mathrm{B}T)^{-1}$, $k_\mathrm{B}$ being Boltzmann's constant and T the absolute temperature \cite{QuOpticsBarnett}. The noise is still smaller than for the squeezed vacuum, but significantly bigger than for a coherent state.\newline
For a displaced thermal state, we obtain
\begin{eqnarray}
\frac{\langle \delta\hat{W}^2\rangle_\mathrm{DisplThermal}}{\langle \delta\hat{W}^2\rangle_\mathrm{Coh}} = \frac{D_{00}^2}{F_{00}} \left(2 - \frac{\bar{n}_\mathrm{th}}{\bar{n}_0}\right)\bar{n}_\mathrm{th} +1,
\end{eqnarray}
with $\bar{n}_0 = \bar{n}_\mathrm{th} + |\alpha_0|^2$, where $|\alpha_0|$ is the displacement \cite{VogelQuOptics}.\newline
Fig.\,\ref{fig:SingleMode}(a) illustrates the dependence of the beam width noise on the quantum state for the example of a fundamental Gaussian mode: the relative beam width uncertainty is plotted against the mean photon number for the quantum states discussed above. The coherent state which serves as a reference is thus set to 1. The vacuum squeezed state and the thermal state exhibit a relative beam width noise higher than 1 that is increasing linearly with the photon number. Furthermore, we plotted the noise for a displaced thermal state with two thermal photons. If more photons are added, they contribute to the displacement. From the figure, one can tell that in this case, the noise is smaller than for a non-displaced thermal state. For the Fock state, the relative noise is constantly equal to $0.5$. The beam width noise is thus reduced compared to the coherent state. The same is valid also for an amplitude squeezed state with a quantum noise reduction of -3\,dB.\newline
Fig.~\ref{fig:SingleMode}(b) illustrates an example of the dependence of the beam width noise on the spatial mode. The beam width noise of a coherent state in a Laguerre-Gauss mode LG$_{lp}$ is plotted for the radial index $p=0$ and different azimuthal indices $l$. The squared mean value of the beam width is used for normalization. Here, we are thus investigating the quantity $\frac{\langle\delta\hat{W}^2\rangle}{\langle \hat{W}\rangle^2}$ in order to learn about the influence of the mode's purely spatial properties on the beam width noise. For the sake of simplicity, we chose $\bar{n}_0 = 0$ for the plot in Fig.~\ref{fig:SingleMode}(b). With increasing $l$, the size of the beam's cross section increases, in particular compared to the fundamental mode. It seems plausible that the quantum noise is getting less significant or influential for a larger total beam size. 
Another type of mode that is particularly interesting in this context is a flattened Gaussian beam \cite{FlattenedGaussian}. It has sharp edges and the spatial distribution looks similar to a top hat function. For such a flattened Gaussian beam as shown in Fig.\,\ref{fig:DetMode}(b), we determine $\frac{\langle\delta\hat{W}^2\rangle_\mathrm{Flat}}{\langle\delta\hat{W}^2\rangle_\mathrm{Coh}} = 0.27$ and $\frac{\langle\delta\hat{W}^2\rangle_\mathrm{Flat}}{\langle \hat{W}\rangle^2_\mathrm{Flat}} = 0.36$ for a Fock state. It is thus in a similar range as the higher order LG-modes, but significantly smaller than the one for a fundamental Gaussian beam. Our examples for LG and Flattened Gaussian modes show that the beam width noise depends on the classical mode.\newline
We have shown that it is possible to reduce the quantum uncertainty in the beam width by an appropriate choice of the spatial mode and the quantum state. Nevertheless, in none of our examples, the noise was minimized to zero. The reason for this is that we are facing a multimode problem: the uncertainty in the beam width depends not only on the uncertainty in the mean field mode, but also on the uncertainty in the residual modes. Changing the quantum properties of the mean field mode can thus not be sufficient. This will become clear in the following section.
\subsection{Multimode States with Small Quantum Noise} \label{Sec:SmallQFluc}
For our investigation of the beam width noise of multimode states, we assume small quantum noise. We may thus use the linearization $\delta \hat{a}_i = \hat{a}_i - \langle \hat{a}_i \rangle$, with $\delta \hat{a}_i\ll \langle \hat{a}_i \rangle$. Compared to the general case discussed in the very beginning of Sec.\,\ref{SecGenBWN}, the formulas simplify significantly. For the deviation from the mean value, we yield the relatively simple expression:
\begin{eqnarray}
\delta \hat{W}= \frac{1}{N_\mathrm{all}}\sum_{ij} D_{ij} (\langle \hat{a}_i^\dagger \rangle \delta\hat{a}_j + \delta\hat{a}_i^\dagger\langle\hat{a}_j\rangle).
\end{eqnarray}
We construct a basis $\{u_i(x,y)\}$ such that the mean value of the electric field is nonzero only in the first mode $u_0$ \cite{Treps2005} called mean field mode. Accordingly, we can then write $\langle \hat{a}_i \rangle = \langle \hat{a}_0 \rangle \delta_{i0}$, and by choosing the phase appropriately we can assume further $\langle \hat{a}_0  \rangle = \langle \hat{a}_0^\dagger \rangle$. Thanks to these relations, we obtain 
\begin{eqnarray}
\langle\delta\hat{W}^2\rangle = \langle \hat{a}_0 \rangle^2 \frac{F_{00}}{N_\mathrm{all}^2} \langle(\delta\hat{A}+\delta\hat{A}^\dagger)^2\rangle,
\end{eqnarray}
where $\hat{A} = \frac{1}{\sqrt{F_{00}}} \sum_i D_{0i} \hat{a}_i$ and $\hat{A}^\dagger = \frac{1}{\sqrt{F_{00}}} \sum_i D_{i0} \hat{a}_i^\dagger$ can be regarded as the annihilation and creation operators of a new mode that we choose to call detection mode $v_0$. The spatial shape of $v_0$ will be discussed in the next section.
 The noise in the beam width is completely attributed to noise in the amplitude quadrature $\hat{X}_{w_0}^{(+)} = \hat{A}^\dagger+\hat{A}$ of this detection mode:
\begin{eqnarray}
\langle\delta\hat{W}^2\rangle = \langle \hat{a}_0 \rangle^2 \frac{F_{00}}{N_\mathrm{all}^2} \langle\delta^2 \hat{X}_{v_0}^{(+)}\rangle \label{EqLinVarW}
\end{eqnarray}
This also means that by squeezing the amplitude quadrature $\hat{X}_{v_0}^{(+)}$ of the detection mode, we can achieve a reduction of the quantum noise in the beam width.\newline
The detection mode for the beam width depends solely on the field under investigation, it refers to a fundamental property of the beam itself. In the case of the quantum laser pointer described in \cite{QuLaserPointer}, in contrast, the detection mode refers to the given problem of a stabilization of the beam's position with respect to a reference frame.
\subsubsection{The Detection Mode} \label{SecDetMod}
A practical approach for how to achieve a reduction of the noise in a beam's width is thus to mix the bright beam defining the mean field mode with an amplitude squeezed state in the associated detection mode. Also with regard to this, it is interesting to study the nature of the detection mode further, and, in particular, to determine its spatial distribution. The annihilation operator $\hat{A}$ defines the detection mode, we can deduce the spatial shape of $v_0$ from it as \cite{bookClaude} and determine
\begin{eqnarray}
v_0(x,y) = \frac{1}{\sqrt{F_{00}}}(x^2+y^2) u_0(x,y).
\end{eqnarray}
\begin{figure}[h]%
    \subfloat[][]{\includegraphics[width=0.5\textwidth]{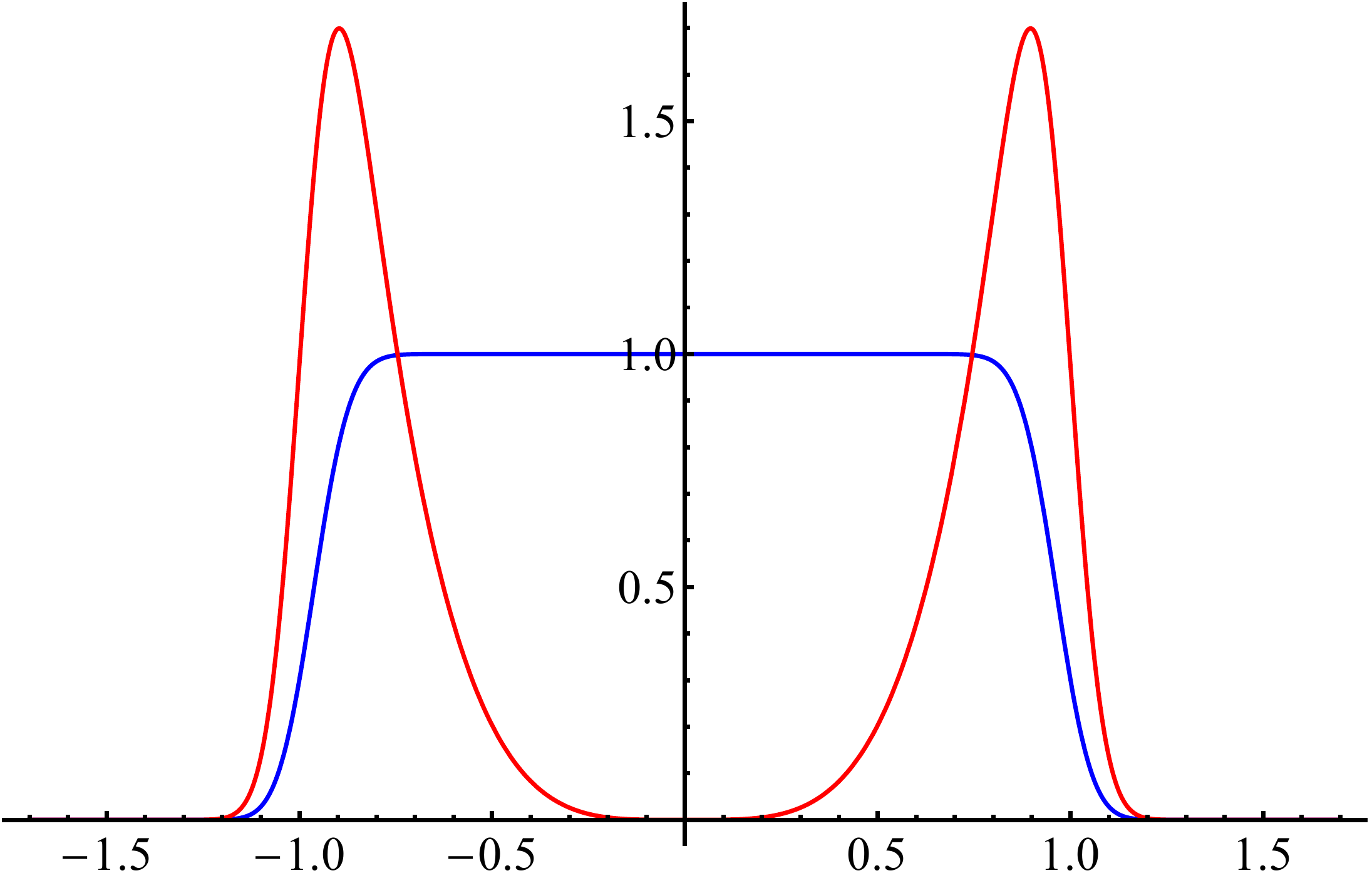}}
   \hfill
   \subfloat[][]{\includegraphics[width=0.5\textwidth]{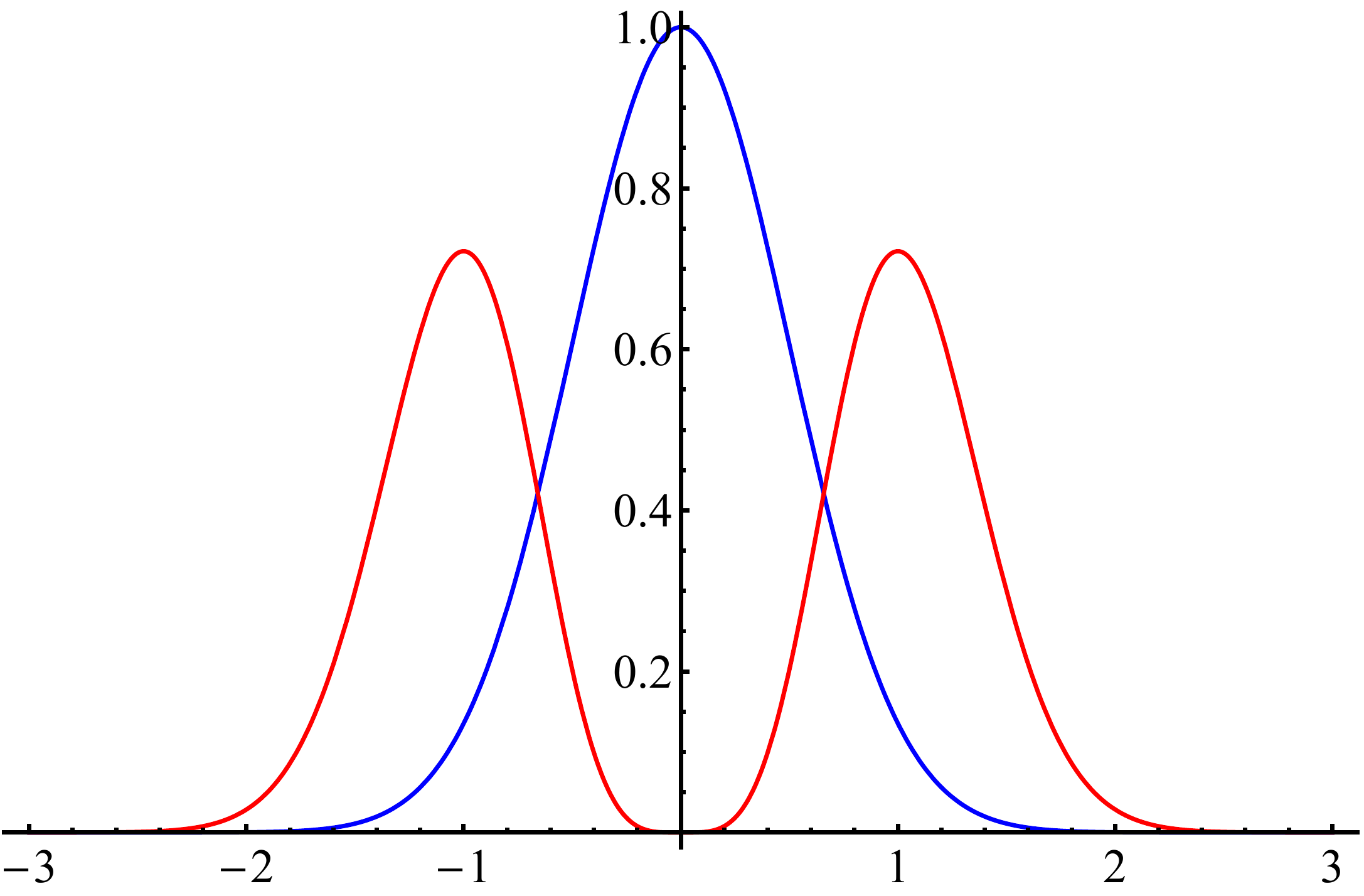}}
    \caption{The plots show two examples of mean field modes and their respective detection mode: In (a), the mean field mode (blue) is a fundamental Gaussian mode, and its detection mode (red) a superposition of Hermite-Gauss modes of the zeroth and the second order. (b) shows a flattened Gaussian beam of the order $N=$30  as the mean field mode and its detection mode consisting of two peaks at the position of the edges of the flattened Gaussian beam.}
    \label{fig:DetMode}
\end{figure}
As one call tell from this expression, the detection mode is not orthogonal to the mean field mode $u_0$. Consequently, the noise in the beam width is not independent from the noise in the mean field mode, and is also influenced by the noise in other modes of the initial basis. This fits to what we have seen in the previous section for the single mode states: we were not able to reduce the noise in the beam width to zero by solely changing the quantum state of the mean field mode. 
We may construct an orthonormal basis with the first mode being the detection mode $v_0$. 
The residual part of the mean field shall be contained in the second mode $v_1$. The mean field mode $u_0$ can then be expressed as a superposition of the detection mode $v_0$ and $v_1$:
\begin{eqnarray}
u_0(x,y) = \frac{D_{00}}{\sqrt{F_{00}}} v_0(x,y) + \sqrt{1-\frac{D_{00}^2}{F_{00}}} v_1(x,y).
\end{eqnarray}
To develop a better understanding of the nature of the detection mode, we derive it explicitly for two examples of the mean field mode: the one of a fundamental Gaussian mode, and the one of a Flattened Gaussian Beam. 
For the mean field mode being a fundamental Gaussian mode, the detection mode is a linear combination of Hermite-Gauss modes of the zeroth order $u_{\mathrm{HG}_0}$ and of the second order $u_{\mathrm{HG}_2}$:
\begin{eqnarray}
v_0(x,z) = \sqrt{\frac{2}{3}}u_{\mathrm{HG}_2}(x,z) + \sqrt{\frac{1}{3}} u_{\mathrm{HG}_0}(x,z)
\end{eqnarray}
For the sake of simplicity and clarity, we express it in only one transverse dimension. In Fig.~\ref{fig:DetMode}(a), the fundamental Gaussian mode and its detection mode are illustrated: the detection mode has two peaks at the position of the edges of the fundamental HG-mode. Its shape is very intuitive, if one keeps in mind that the amplitude noise of this mode is responsible for the noise in the beam width. If one increases the intensity at the edges of the beam, its width increases. Thus, if the detection mode consisting of two peaks right at the edges of the mean field is given an amplitude, the width of the beam will increase. Amplitude noise in this mode will thus cause beam width noise.\newline
Fig.~\ref{fig:DetMode}(b) shows a Flattened Gaussian Beam and its detection mode. Similarly to the case of a fundamental HG-mode, the detection mode consists of two peaks at the position of the edges of the mean field mode. Here, the edges are much sharper, and, accordingly, the peaks of the detection mode are also significantly narrower.\newline
\section{The Conjugate Counterpart of the Beam Width}
Another question that naturally arises is the one of a conjugate variable of the beam width. Due to the definition of our beam width operator $\hat{W}$, a canonically conjugate $\hat{C}$ with $[\hat{W},\hat{C}] = \mathrm{const}$ cannot exist. This is because with the product $\hat{a}^\dagger\hat{a}$ in $\hat{W}$, the commutator will never give a constant, but always an operator. For the linearized approach, one can find a canonically conjugate via the detection mode $v_0$, as the conjugate variable is then associated with $-i v_0$. As the physical interpretation of that variable is not obvious, we did not pursue this any further.\newline
Nevertheless, we may find the Fourier conjugate $c$ of the beam width $w$, which is its Fourier transform and fulfills $\Delta w \Delta c \sim 1$. The angular spread satisfies this; through conclusion of analogy one can derive the conjugate of $\hat{W}$ from classical relations \cite{Lin}. While the beam width is defined via the variance in the spatial domain, i.e.  $\iint u_i^*(x,y) (x^2 + y^2) u_j(x,y) \mathrm{d}x\mathrm{d}y$, the angular spread is then given as its equivalent in Fourier space: $\iint u_i^*(x,y) (p_x^2 + p_y^2) u_j(x,y) \mathrm{d}x\mathrm{d}y$. Using $p_x \leftrightarrow \frac{1}{i k} \partial_x$ and $p_y \leftrightarrow \frac{1}{i k} \partial_y$, we yield the operator $\hat{\Theta}$:
\begin{eqnarray}
\hat{\Theta} &= \frac{1}{N_\mathrm{all}} \sum_{ij} \tilde{D}_{ij} \hat{a}_i^\dagger \hat{a}_j,
\end{eqnarray}
with
\begin{eqnarray}
\tilde{D}_{ij} &= - \frac{1}{k^2} \iint u_i^*(x,y) (\partial_x^2 + \partial_y^2) u_j(x,y) \mathrm{d}x\mathrm{d}y.
\end{eqnarray}
The calculations presented above for $\hat{W}$ can be performed in complete analogy for $\hat{\Theta}$. The results differ only in the integral expressions $\tilde{D}_{ij}$ and \mbox{$\tilde{F}_{il} = \frac{1}{k^4} \iint u_i^*(x,y) (\partial_x^2 + \partial_y^2)^2 u_l(x,y) \mathrm{d}x\mathrm{d}y$}. For the multimode case, we may again apply the linearization of the annihilation and creation operators assuming small quantum noise. We then find a detection mode \mbox{$m_0(x,y) = - \frac{1}{k^2 \sqrt{\tilde{F}_{00}}} (\partial_x^2+\partial_y^2)u_0(x,y)$}. If the mean field is in a fundamental Gaussian mode, we can express it again as a superposition of the fundamental and the second order HG modes as \mbox{$m_0(x) = \sqrt{\frac{1}{3}} u_{\mathrm{HG}_0}(x)- \sqrt{\frac{2}{3}}u_{\mathrm{HG}_{2}} (x)$}.\newline
It is meaningful in particular in the Fourier space, where \mbox{$\tilde{m}_0(p_x,p_y) \propto (p_x^2+p_y^2)\tilde{u}_0(p_x,p_y)$}. The analogy to the detection mode for the beam width becomes obvious: the angular aperture detection mode has the same shape in the Fourier space as the beam width detection mode in space.\newline
\section{Conclusion and Outlook}
In this article, we provide a study of the quantum noise in the width of a paraxial beam. We investigate the influence of different spatial modes and quantum states on the noise in the beam width for single mode states. We show ways to reduce the noise by an appropriate choice, i.e. for example by using Fock or amplitude squeezed states. For multimode states, we present the so-called detection mode, the amplitude noise of which affects the noise in the beam width. One of the possible methods to minimize the beam width noise is thus to squeeze the amplitude quadrature of that mode. The Fourier conjugate of the beam width is found to be the angular spread of the light beam. While all modes discussed in this manuscript have been scalar, the formulas may be applied for vectorial modes, i.e., modes with non-uniform polarization patterns, as well. The overlap integrals $D_{ij}$ and $F_{il}$ then contain the scalar product of the vectorial modes and, for multimode states, the polarization structure may indeed have an influence on the noise in the beam width.\newline
In a next step, we plan to extend our studies to the case of strong focussing. Furthermore, an experimental proof of our theoretical predictions is envisaged.

\section*{References}
\bibliographystyle{iopart-num}

\end{document}